# Accelerated Nuclear Magnetic Resonance Spectroscopy with Deep Learning

Xiaobo Qu[1*], Yihui Huang[1], Hengfa Lu[1], Tianyu Qiu[1], Di Guo[2], Tatiana Agback[3], Vladislav Orekhov[4], Zhong Chen[1*]

**Nuclear magnetic resonance (NMR) spectroscopy serves as an indispensable tool in chemistry and biology but often suffers from long experimental time. We present a proof-of-concept of application of deep learning and neural network for high-quality, reliable, and very fast NMR spectra reconstruction from limited experimental data. We show that the neural network training can be achieved using solely synthetic NMR signal, which lifts the prohibiting demand for large volume of realistic training data usually required in the deep learning approach.**

Nuclear magnetic resonance (NMR) spectroscopy is an invaluable biophysical tool in modern chemistry and life sciences. Examples include characterization of complex protein structures[1, 2] and studies disordered[3] and short-lived molecular systems[4]. However, duration of NMR experiments increase rapidly with spectral resolution and dimensionality[5, 6], which often imposes unbearable limitations due to low sample stability and/or excessive costs of NMR measurement time. To accelerate the data acquisition and optimize sensitivity, modern NMR experiments are often acquired using the Non-Uniform Sampling (NUS) approach, where only a small fraction of traditional NMR measurements, usually called free induction decay (FID), is performed and, thus, only a fraction of measurement time is spent.

Over the past two decades, several methods [5, 7, 8, 9, 10, 11, 12] have been established in the NMR field to reconstruct high quality spectra from NUS data. In all case, a prior knowledge or assumption are incorporated in order to compensate for missing information introduced by the NUS scheme. Examples include the maximum entropy[6], spectrum sparsity in compressed sensing[9, 10, 13, 14], spectral line-shape estimation in SMILE[15], tensor structures in MDD[5] or Hankel tensors[11], and exponential nature of NMR signal in low rank[7]. Thus, although spectra are reconstructed well with these approaches, a number of important practical limitations and conceptual question remain. Thus, despite of varying implementations, algorithms of all these methods are iterative and require lengthy calculations and/or use of super-computers. Pros and cons of applying different prior assumptions are not well understood and combination of the best features, while avoiding the negative sides of different approaches is problematic.

Motivated by the exciting achievements of deep learning (DL)[16, 17], a representative artificial intelligence using neural networks, we will explore the end-to-end mapping with DL for the NMR spectra reconstruction, enabling fast and high-quality reconstructions. In contrast to the traditional methods that take advantage of one or more predefined priors for reconstruction, for instance, sparsity and low rank, the proposed DL approach mines the underlying information embedded in data and thus does not require any predefined priors.

A critical challenge of the DL is that it requires an enormous amount of realistic experimental data at the training stage. Whilst obtaining of such a gigantic data set is practically impossible due to NMR sample and instrument time limitations, our work demonstrates that successful training of the neural network in the DL is possible using solely synthetic data. These are generated using the classic assumption that NMR FID is a superposition of small number of exponential functions[6, 7]. The strategy of using synthetic data for training is beyond the traditional DL approach that requires huge volume of practical data. This work suggests a way for bridging the traditional signal modeling to DL and for enabling smart artificial intelligence computational tools in applications that lack enough practical data to train the neural network. This work can be treated as a proof-of-concept for DL NMR spectroscopy.

Reconstructing a spectrum from NUS data is equivalent to mapping of the input undersampled FID signal to the target spectrum. In the DL NMR, a neural network is trained to perform the mapping as shown in Figure 1. First, the spectrum artifacts introduced by NUS are removed with dense convolutional neural network (CNN) and then intermediately reconstructed spectra are further refined to maintain the data consistency to the sampled signal. Artifacts are gradually removed as the stage of reconstruction increases and the final spectrum is produced after several stages. In this implementation, dense CNN is chosen because it ensures maximum information flow between layers in the neural network[18] while data consistency constraint the reconstruction subjecting to the sampled data points[19, 20].

[1]Department of Electronic Science, Fujian Provincial Key Laboratory of Plasma and Magnetic Resonance, State Key Laboratory of Physical Chemistry of Solid Surfaces, Xiamen University, Xiamen 361005, China. [2]School of Computer and Information Engineering, Xiamen University of Technology, Xiamen 361024, China. [3]Department of Molecular Sciences, Swedish University of Agricultural Sciences, Uppsala, Sweden. [4]Department of Chemistry and Molecular Biology, University of Gothenburg, Box 465, Gothenburg 40530, Sweden.
Correspondence should be addressed to Xiaobo Qu (quxiaobo@xmu.edu.cn) or Zhong Chen (chenz@xmu.edu.cn).

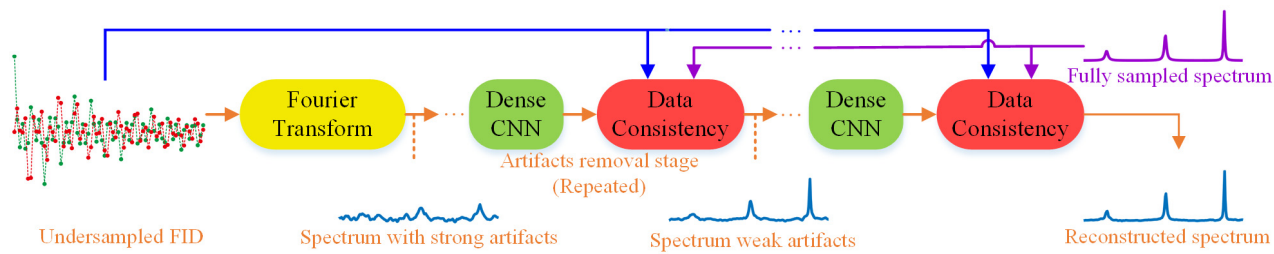

Figure 1. Flowchart of deep learning NMR spectroscopy. Note: Please refer to Supplement S1 for more details.

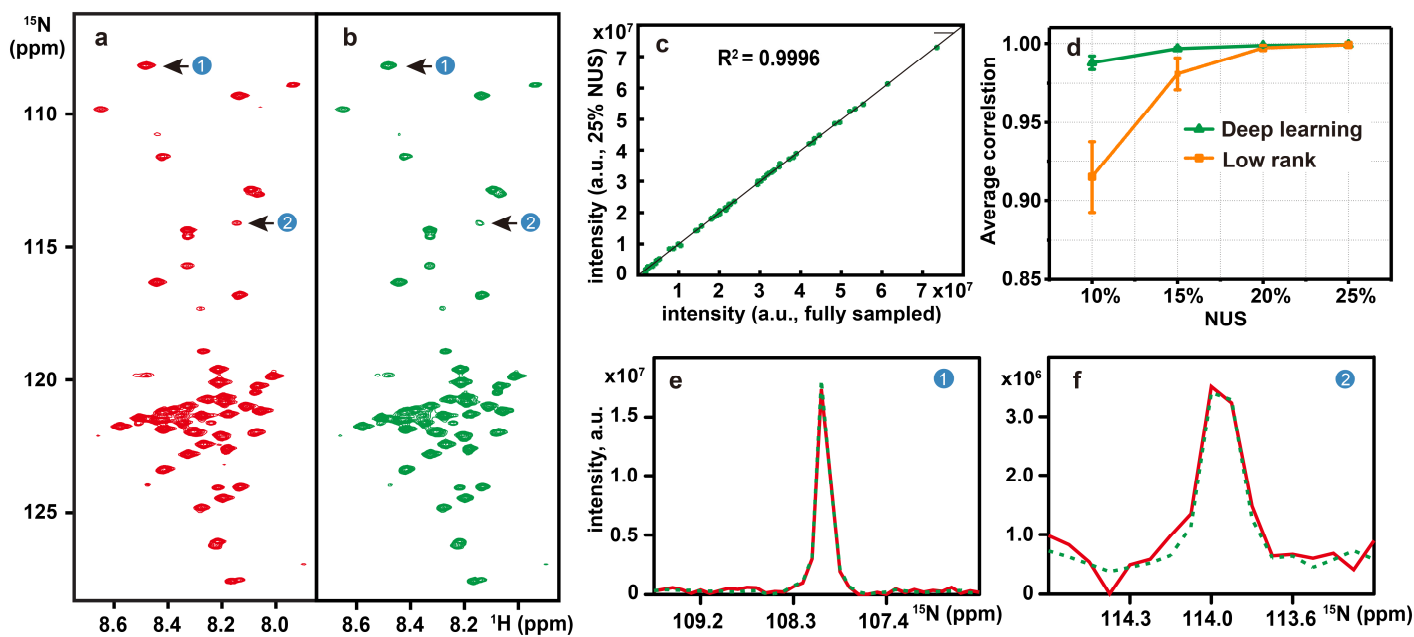

Figure 2. Reconstruction of a 2D $^1$H–$^{15}$N HSQC spectrum of the cytosolic domain of CD79b protein from the B-cell receptor. (a) and (b) are the fully sampled spectrum and deep learning NMR reconstruction from 25% NUS data, respectively. (c) Peak intensity correlations between fully sampled spectrum and reconstructed spectrum. (d) denotes the peak intensity correlation obtained with the deep learning and low rank methods under different NUS levels. (e) and (f) are zoomed out 1D $^{15}$N traces of (a). Red and green lines represent the reference and the reconstructed spectra, respectively. Note: The $R^2$ denotes the square of Pearson correlation coefficient. The closer the value of $R^2$ gets to 1, the stronger the correlation between the reference and the reconstructed spectra is. The average and standard deviations of correlations in (d) are computed over 100 NUS trials.

The key issue for DL NMR is to learn the mapping. We used computer to generate the fully-sampled time domain NMR signal, from which undersampled NUS signal was obtained using Poisson gap sampling scheme (See Supplement S1.1 for more details). Given the synthetic NUS signal $y$ and the corresponding target spectrum $s$ produced from the fully sampled time domain data, a large number of pairs $(y_k, s_k)$ ($k$=1, 2, …, $K$) are fed into the neural network to learn the best network parameters $\theta$ that minimizes the least errors $e(\theta) = \sum_{k=1}^{K}(f(y_k, \theta) - s_k)^2$. Therefore, DL provides an optimal mapping $f(y, \theta)$ from the input $y$ to the target spectrum in the sense of least square error for all pairs. Then, for a given undersampled FID $\tilde{y}$ from a NUS experiment, a spectrum $\tilde{s}$ is obtained via $\tilde{s} = f(\tilde{y}, \theta)$.

To demonstrate the applicability of the DL NMR, we first validate the reconstruction performance on several fully sampled 2D and 3D spectra of small proteins. As shown in Figure 2, DL reconstructs excellent 2D $^1$H-$^{15}$N HSQC spectrum from 25% NUS data with correlation of the peak intensity to the fully sampled spectrum reaching 0.9996. Figure 2d indicates that DL is in pair with the state-of-the-art reconstruction techniques[7] in robustness and spectra quality and may even surpass the other methods at low NUS densities (See Supplement S2.2 for more details). High fidelity of the reconstructed peak shapes is illustrated in Figures 2e and 2f. Using the network with same trained parameters, the correlations greater than 0.98 were also obtained for 2D spectra of three other proteins (See Supplement S2.3). High potential of the DL in reconstructing high-quality multi-dimensional spectra is illustrated in Figure 3, exemplified by 3D HNCO for Azurin (14 kDa protein) and 3D HNCACB spectrum for GB1-HttNTQ7 (10 kDa protein). The peak intensity correlations approaching 0.99 for both 3D spectra (Figures 3e and 3f) indicates excellent fidelity of the DL reconstruction.

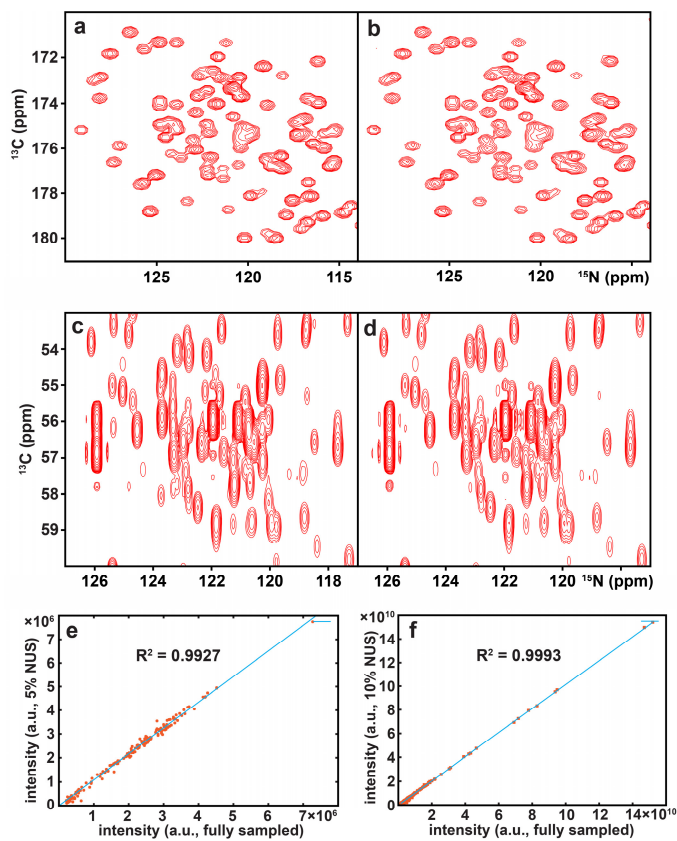

Figure 3. 3D spectra reconstructions for two small proteins, azurin (14 kDa) and GB1-HttNTQ7 (10 kDa). (a-b) Sub-regions of $^{13}C$-$^{15}N$ projections from the fully sampled spectrum of azurin and the deep Learning reconstruction with 5% NUS data. (c-d) Sub-regions of $^{13}C$-$^{15}N$ projections from the fully sampled spectrum of GB1-HttNTQ7 and the deep Learning reconstruction with 10% NUS data. (e-f) Peak intensity correlations between deep learning reconstructions and fully sampled 3D spectra of azurin and GB1-HttNTQ7. Note: The contours of spectra are at the same level.

Figure 4 illustrates quality and performance of DL for challenging cases of a large protein MALT1 (44 kDa ) and an intrinsically disordered protein alpha-synuclein (14.5 kDa) (See Supplement S2.1 for more details). Even with 10% data, DL provides robust high-quality spectra reconstruction. This indicates that DL would enable a considerable acceleration factor, i.e. 10, of the fully sampled data acquisition for these challenging cases.

An important advantage of the DL NMR is fast spectra reconstruction due to harnessing of a non-iterative low-complexity neural network algorithm that allows massive parallelization with graphics processing units. Without compromising the spectra quality (See Supplements S2.2 and S2.3 for detailed comparisons), DL is much faster than other state-of-the-art methods such as low rank[7] and compressed sensing[10]. The comparisons, shown in Figure 5, indicate that the computational time of DL is 4%~8% of that needed by low rank for 2D spectra and 12%~22% of that consumed by compressed sensing for 3D spectra. Although, training of the network is computationally demanding, it is done only once, whereas all the subsequent reconstructions of experimental spectra are fast.

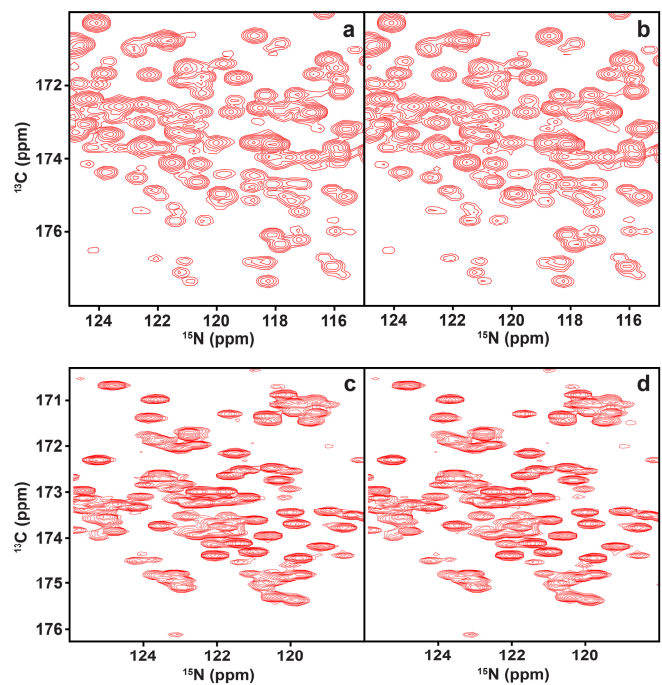

Figure 4. 3D HNCO spectra reconstructions for a large protein, MALT1 (44 kDa), and an intrinsically disordered protein, alpha-synuclein (14.5 kDa). (a-b) Sub-regions of $^{13}C$-$^{15}N$ projections from reconstructions of the MALT1 protein by DL with 30% and 10% NUS data, respectively. (c-d) Sub-regions of $^{13}C$-$^{15}N$ projections from reconstructions of the alpha-synuclein protein by DL with 15% and 10% NUS data, respectively. Note: The experimental data for MALT1 and alpha-synuclein proteins were acquired under 30% and 15% NUS, respectively. Further randomly under sampling is retrospectively applied to the experimental NUS data to emulate sampling at lower NUS densities. The contours in the pairs of spectra are at the same level.

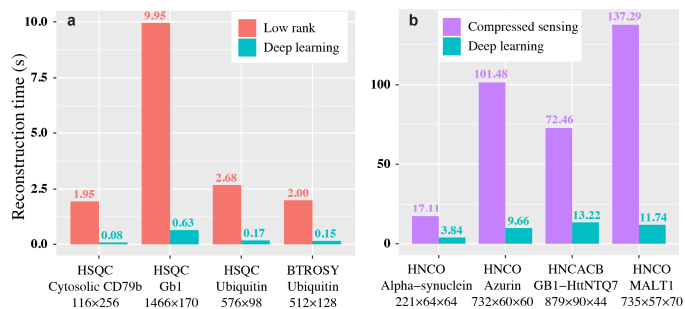

Figure 5. Computational time for the reconstructions of (a) 2D spectra and (b) 3D spectra. The spectra type, its corresponding protein and spectra size after routine processing of the direct dimension are listed below each bar. Details about the time comparisons are found in Supplement S3.

In summary, we present the proof-of-concept demonstration of application the DL for reconstructing high quality NMR spectra of small, large and disordered proteins from NUS data. This result opens an avenue for application of DL and possibly other artificial intelligence techniques in biological NMR. Not limited to NMR, we demonstrated that DL can be achieved using purely synthetic training sets. Thus, the exponential function reconstruction may also be valuable to other biomedical imaging tools[21, 22]. Another important feature of the DL is its inherent ability to mine underlying properties of the signal, which may give the DL NMR the upper hand in crucial applications, where it is hard to define a good model for the signal of interest.

## Acknowledgements

Authors thank Marius Clore and Samuel Kotler for providing the 3D HNCACB data; Jinfa Ying for assisting processing and helpful discussion on the 3D HNCACB spectrum; Luke Arbogast and Frank Delaglio for providing the 2D HSQC spectrum of GB1; Esmeralda Woestenenk for help with MALT1 protein production. This work was supported in part by the National Natural Science Foundation of China (NSFC) under grants 61571380, 61871341 and U1632274, the Joint NSFC-Swedish Foundation for International Cooperation in Research and Higher Education (STINT) under grant 61811530021, the National Key R&D Program of China under grant 2017YFC0108703, the Natural Science Foundation of Fujian Province of China under grant 2018J06018, the Fundamental Research Funds for the Central Universities under grant 20720180056, the Swedish Research Council under grant 2015–04614 and the Swedish Foundation for Strategic Research under grant ITM17-0218.

## Author contributions

X. Qu conceived the idea and designed the study, X. Qu, V. Orekhov and Z. Chen supervised the project, H. Lu and Y. Huang implemented the method and produced the results, H. Lu, Y. Huang and T. Qiu drew all the figures for the manuscript. T. Agback acquired the MALT1 protein data. All authors were involved in the data analysis. The manuscript was drafted by X. Qu and improved by X. Qu, H. Lu, Y. Huang, D. Guo and V. Orekhov. X. Qu, D. Guo and Z. Chen acquired research funds and provided all the needed resources.

## Competing interests

The authors declare no competing interests.

## Data availability

The synthetic FID data used for training are shared at http://csrc.xmu.edu.cn. The 2D HSQC spectrum of Gb1 is available from NMRPipe website (https://www.ibbr.umd.edu/nmrpipe/index.html). The 3D HNCO spectrum of Azurin is available from MddNMR website (http://mddnmr.spektrino.com/). The spectra data of 2D HSQC of Cytosolic CD79b, 2D HSQC of Ubiquitin, 2D BTROSY of Ubiquitin, 3D HNCO of Alpha-synuclein, 3D HNCO of MALT1 and 3D HNCACB of GB1-HttNTQ7 are available from the corresponding author upon reasonable request.

## Code availability

Code is available from the corresponding author upon reasonable request.

# Supplement S1: Methodology

In the following, we first illustrate the detailed architectures (Fig. S1-1) of the deep learning (DL) NMR and then explain each processing parts separately, following the processing of data flow.

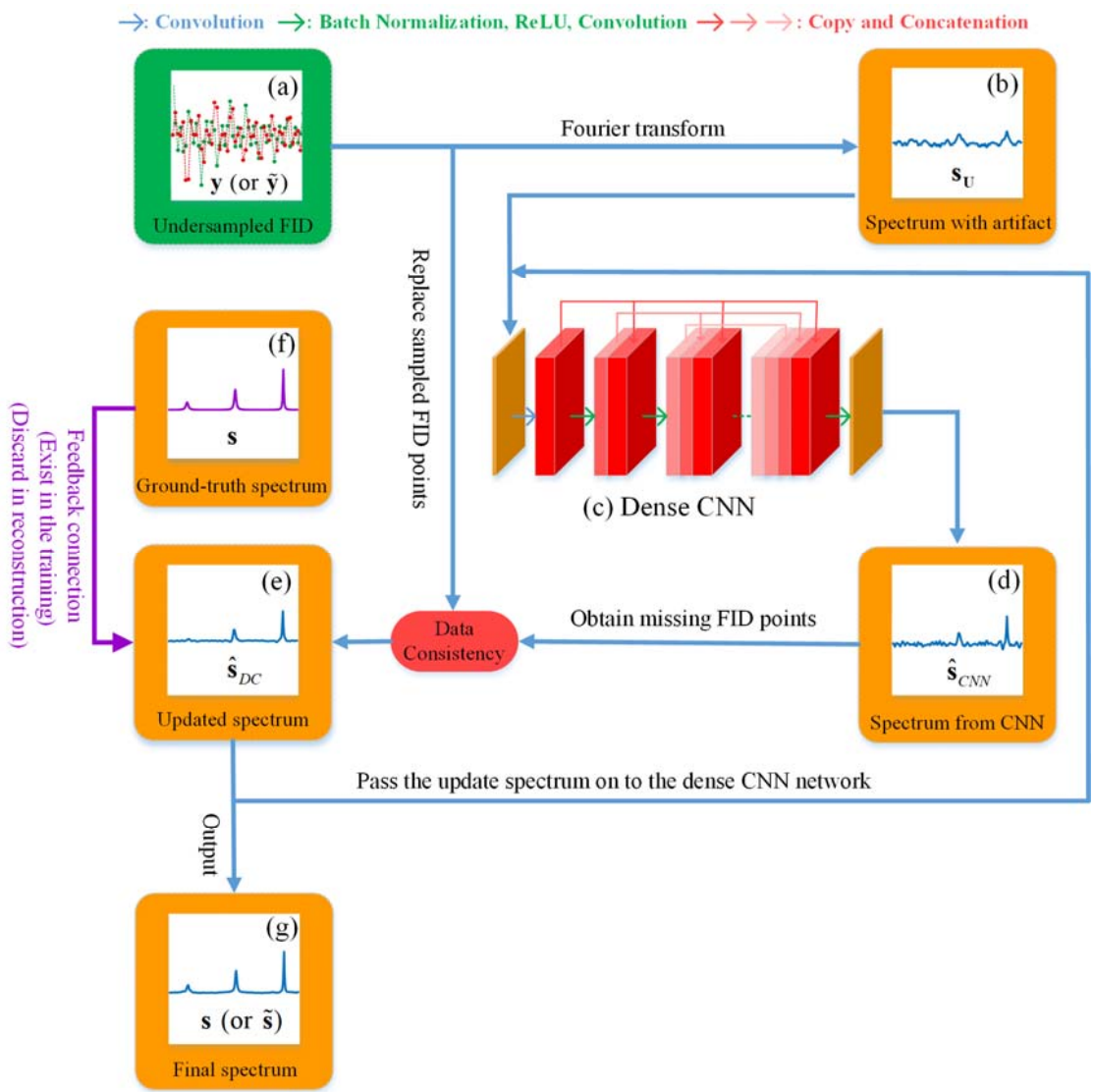

Figure S1-1. The detailed architectures of DL NMR. (a) The undersampled FID, (b) the spectrum with strong artifacts, (c) dense CNN, (d) the output of dense CNN, (e) the updated spectrum from data consistency, (f) fully sampled spectrum, (g) the output of the whole network. Note: The green and orange blocks denote signals in time and frequency domains, respectively. Signals or modules that are marked with the purple color are only existed in the training phase.

The implementation of DL NMR includes two phases, training phase and prediction phase. In the training phase, with the computer-simulated undersampled FID $\mathbf{y}$ to the target spectrum $\mathbf{s}$, a large number of FID/spectrum pairs $(\mathbf{y}_k, \mathbf{s}_k)(k=1,2,\cdots,K)$ are input into the neural network to learn the best network parameters $\hat{\boldsymbol{\theta}}$. In the reconstruction phase, given an undersampled FID $\tilde{\mathbf{y}}$ acquired in the NUS experiment, a spectrum $\tilde{\mathbf{s}}$ is obtained via $\tilde{\mathbf{s}} = f(\tilde{\mathbf{y}}, \hat{\boldsymbol{\theta}})$ where $f$ is the trained mapping from undersampled FIDs to spectra. Both training and reconstruction phases are detailed illustrated in the following.

## 1.1 Training phase

### 1.1.1 Generate the fully sampled spectrum and the undersampled FID

Our method solely uses the synthetic data as training data, which is significantly different from many deep learning approaches that utilize the realistic data as training data. The fully sampled spectrum satisfies $\mathbf{s}=\mathbf{Fx}$, where $\mathbf{F}$ is the Fourier transform and $\mathbf{x}$ is the fully sampled FID, and the undersampled FID obeys $\mathbf{y}=\mathbf{Ux}$, where $\mathbf{U}$ is the undersampling operator, are generated as follows:

The fully sampled FID $\mathbf{x}$ is simulated according to the classical exponential function modeling as[1-5]:

$$x_n = \sum_{j=1}^{J} \left( A_j e^{i\phi_j} \right) e^{-\frac{n\Delta t}{\tau_j}} e^{in\Delta t 2\pi\omega_j}, \quad \text{(S1-1)}$$

where $J$ is the number of exponentials, $A_j$, $\emptyset_j$, $\tau_j$ and $f_j$ are the amplitude, phase, decay time and frequency, respectively, of the $j^{th}$ exponential. By varying these parameters according to Table S1-1, there are 40000 FIDs are simulated.

For each fully sampled FID $\mathbf{x}$, a corresponding under-sampling operator $\mathbf{U}$ is generated following the Poisson-gap sampling scheme[8]. Let $k$ denote the $k^{th}$ sampling trial, then multiple pairs of $(\mathbf{y}_k, \mathbf{s}_k)$ $(k=1,2,\cdots,K)$, composed of the undersampled FID $\mathbf{y}_k$ and the fully sampled spectrum $\mathbf{s}_k$, are formed and used to train the neural network. In this work, we simulate $K=40000$ pairs.

Table S1-1. Parameters for 1D synthetic FID

| Parameters | Number of Peaks ($J$) | Amplitude ($A$) | Frequency ($\omega$) | Decay time ($\tau$) | Phase ($\emptyset$) |
|---|---|---|---|---|---|
| Minimum | 1 | 0.05 | 0.01 | 10 | 0 |
| Increment | 1 | continuous | continuous | continuous | continuous |
| Maximum | 10 | 1 | 0.99 | 179.2 | 2π |

Note: The FID is normalized so that the maximal magnitude of each spectrum is 1.

### 1.1.2 Generate the initial spectrum from the undersampled FID

The initial spectrum that inputs the neural network is computed as $\mathbf{s}_U = \mathbf{F}^H \mathbf{U}^T \mathbf{y}$, where $\mathbf{U}^T$ is the adjoint operator of $\mathbf{U}$ and $\mathbf{F}^H$ is the forward Fourier transform. This initial spectrum is with strong artifacts since those unsampled FID data are filled with zeros on non-acquired positions.

Since an undersampled FID $\mathbf{y}_k$ corresponds to one NUS sampling $\mathbf{U}_k$, thus the generated initial spectrum will be $\mathbf{s}_{U_k} = \mathbf{F}^H \mathbf{U}_k^T \mathbf{y}_k (k=1,2,\cdots,K)$ and $K=40000$ in the implementation.

### 1.1.3 Reduce spectrum artifacts with dense neural network

The spectrum $\mathbf{s}_U$ is fed into the densely connected convolutional neural networks (Fig.S1-1(c)), known as dense CNN [6]. This neural network learns a map $f_{CNN}$ to reduce the spectrum artifact and yield the 'clean' spectrum denoted as $\hat{\mathbf{s}}_{CNN}$.

The structures of dense CNN (Fig. S1-1(c)) include 8 convolutional layers. Between adjacent layers of dense CNN, there exists the batch normalization followed by the ReLU activation function. With the initial spectrum as input, first convolutional layer produces 16 spectra while the rest of convolutional layers each output 12 spectra except for the last layer which provides only one spectrum - the spectrum $\hat{\mathbf{s}}_{CNN}$. The $l^{th} (2 \leq l \leq 8)$ layer takes outputs of preceding $(l-1)^{th}$ layers, i.e., $16+12\times(l-2)$ spectra.

### 1.1.4 Enforce the spectrum to maintain data consistency

A data consistency module is incorporated to ensure reconstructed spectra are aligned to acquired data. Given the output of dense CNN $\hat{\mathbf{s}}_{CNN}$, the spectrum is modified as

$$\hat{\mathbf{s}}_{DC} = \arg\min_{\mathbf{s}_{DC}} \left\{ \left\| \mathbf{s}_{DC} - \hat{\mathbf{s}}_{CNN} \right\|^2 + \lambda \left\| \mathbf{y} - \mathbf{U}\mathbf{F}^T \mathbf{s}_{DC} \right\|^2 \right\}, \qquad (S1\text{-}2)$$

where $\|\cdot\|$ denotes the norm of a vector, $\mathbf{F}^T$ the inverse Fourier transform, $\mathbf{s}_{DC}$ the underlying spectrum to be optimized, and $\hat{\mathbf{s}}_{DC}$ is the output of data consistency module. A closed form solution of Eq. (S1-2) is

$$\hat{\mathbf{s}}_{DC} = \mathbf{F}\left( \lambda \mathbf{U}^T \mathbf{U} + \mathbf{1} \right)^{-1} \left( \lambda \mathbf{U}^T \mathbf{y} + \mathbf{F}^T \hat{\mathbf{s}}_{CNN} \right), \qquad (S1\text{-}3)$$

where $\mathbf{1}$ is an identity matrix and $(\cdot)^{-1}$ denotes the inverse of a matrix. Let the FID of $\hat{\mathbf{s}}_g$ be $\hat{\mathbf{x}}_g = \mathbf{F}^T \hat{\mathbf{s}}_g$, then Eq. (S1-3) is equivalent to the following relationship

$$\left( \hat{\mathbf{x}}_{DC} \right)_n = \begin{cases} \left( \mathbf{F}\hat{\mathbf{s}}_{CNN} \right)_n, & n \notin \Omega \\ \dfrac{\left( \mathbf{F}\hat{\mathbf{s}}_{CNN} \right)_n + \lambda \mathbf{y}_n}{1 + \lambda}, & n \in \Omega \end{cases}, \qquad (S1\text{-}4)$$

where $\Omega$ is the set of positions for sampled FID and $n$ is the index of the FID. The Eq. (S1-4) implies that the FID at the location of sampled data points should be balanced between the acquired data points in the initial data $\mathbf{y}$ and the predicted data point obtained with the dense CNN.

For simplicity, we rename the data consistency as a linear function

$$\hat{\mathbf{s}}_{DC} = f_{DC}\left( \hat{\mathbf{s}}_{CNN}, \mathbf{y} \right), \qquad (S1\text{-}5)$$

that maps the input $\left( \hat{\mathbf{s}}_{CNN}, \mathbf{y} \right)$ to the spectrum according to Eq. (S1-3). In our implementation, the produced spectrum of the DC layer was calculated by performing the operations in Eq. (S1-4) on the output of dense CNN with $\lambda = 10^6$ which works well for all the tested spectra.

In this implementation, the two modules described in S1.1.3 and S1.1.4 are combined as one reconstruction stage. As shown in Fig. S1-2, spectrum artifacts (Fig. S1-2(b)) are firstly removed to some degree by the dense CNN (Fig. S1-2(c)) and then the spectrum (Fig. S1-(d)) quality is enhanced by enforcing the data consistency. Further improvement of spectra (Figs. S2-(d)(f)(h)(j)(l)) are retained by repeating reconstruction stage in multiple times.

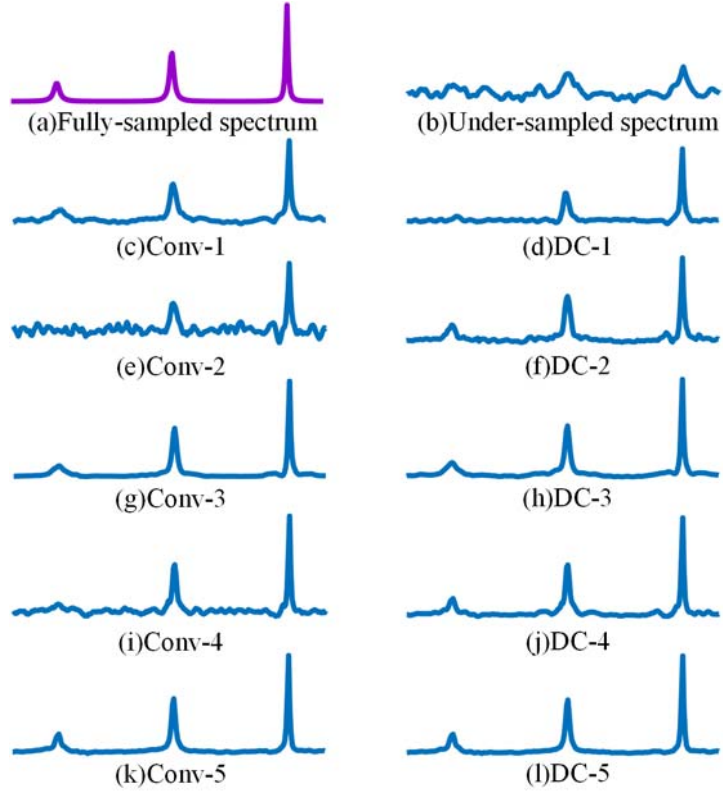

Figure S1-2. The illustrative results of dense CNN and data consistency. Note: *Conv*, *DC*, and the *number (1, 2, 3, 4, 5)* denote the outputs of dense CNN and data consistency, and the order of reconstruction stage, respectively. The final output of the network is (l) DC-5.

1.1.5 Loss function and trained optimal parameters

Let the superscript number $q$ denotes the $q^{th}$ reconstruction stage, then the output at the $q^{th}$ stage is the same as the output of the data consistency module, meaning that $\hat{\mathbf{s}}^q = \hat{\mathbf{s}}^q_{DC}$. The overall loss function in our implementation is

$$\min_{\theta=\{\theta^1,\cdots\theta^q\cdots,\theta^Q\}} \sum_{k=1}^{K} \sum_{q=1}^{Q} \left\| \mathbf{s}_k - \hat{\mathbf{s}}^q_k(\theta^q) \right\|, \tag{S1-6}$$

where $\theta$ is network parameters to be trained, $k$ denotes the $k^{th}$ NUS trial, which is also equal to the number of FIDs. In the implementation, ADAM scheme is adopted to solve Eq. (S1-6)[7]. Therefore, the optimal parameters $\hat{\theta}$ is obtained by minimizing the output of the network for all training data.

## 1.2 Reconstruction phase

In the reconstruction phase, given an undersampled FID $\tilde{\mathbf{y}}$ acquired in the NUS experiment, a spectrum $\tilde{\mathbf{s}}$ is reconstructed according to

$$\tilde{\mathbf{s}} = f(\tilde{\mathbf{y}}, \hat{\theta}), \tag{S1-7}$$

where $f$ is the functions that models the whole processing in the neural network.

One thing should be mentioned is that the feedback connection (purple line in Fig. S1-1) is discarded in the reconstruction since the fully sampled FID is not available in practice.

# Supplement S2: Other Spectra Results

In the following, all non-uniform sampling tables are generated according to Poisson-gap sampling[8].

The proposed deep learning (DL) approach will be compared with two state-of-the-art NMR spectroscopy reconstruction approaches, including low rank (LR)[2] and compressed sensing (CS)[9-11]. In reconstruction of 2D NMR, CS[10] is excluded since the LR[2] has been shown to outperform the CS. Thus, comparing deep learning (DL) with LR is enough to demonstrate the advantage of DL. In the reconstruction of 3D NMR, CS[10] is included but LR[2] is excluded because the former can handle the realistic 3D NMR data while the latter cannot accomplish this yet.

## 2.1 Experiments Setup

The important spectra parameters, including four 2D spectra and four 3D spectra of small, large and intrinsically disordered proteins, are listed in Table S2-1. More details could be found in below experimental descriptions (S2.1.1 and S2.1.2). The direct dimension of all spectra was processed in NMRPipe[12] before performing reconstructions.

Table S2-1. Important parameters of used spectra.

| | Type | Protein | Molecular weight | Spectrometer frequency | Sampling type | References |
|---|---|---|---|---|---|---|
| **2D spectra** | HSQC | Cytosolic CD79b | ~5.7 kDa | 800 MHz | Full | Figure S2-1 |
| | HSQC | Ubiquitin | ~8.6 kDa | 800 MHz | Full | Figure S2-3 |
| | HSQC | Gb1 | ~8.0 kDa | 600 MHz | Full | Figure S2-4 |
| | TROSY | Ubiquitin | ~8.6 kDa | 800 MHz | Full | Figure S2-5 |
| **3D spectra** | HNCO | Azurin | 14 kDa | 800 MHz | Full | Figure S2-7 |
| | HNCACB | GB1-HttNTQ7 | 10 kDa | 700 MHz | Full | Figure S2-10 |
| | HNCO | MALT1 | ~44 kDa | 700 MHz | NUS | Figure S2-13 |
| | HNCO | Alpha-synuclein | ~14.5 kDa | 800 MHz | NUS | Figure S2-15 |

Note: The NUS means that the FID data were acquired on spectrometer in non-uniform sampling mode for reducing data acquisition time.

### 2.1.1 2D Spectra

We used the same 2D $^1$H–$^{15}$N HSQC spectrum (Fig. S2-1) of cytosolic CD79b protein as was described in our previous work[2, 13]. In brief, the spectrum was acquired for 300 μM $^{15}$N-$^{13}$C labeled sample of cytosolic CD79b in 20 mM sodium phosphate buffer, pH 6.7 at 25 °C on 800 MHz Bruker AVANCE III HD spectrometer equipped with 3 mm TCI cryoprobe. The fully sampled spectrum consists of 1024×256 complex points, the direct dimension ($^1$H) has 1024 data points while the indirect dimension ($^{15}$N) 256 data points.

The 2D $^1$H–$^{15}$N HSQC spectrum (Fig. S2-3) of Ubiquitin was acquired from ubiquitin sample at 298.2K temperature on an 800 MHz Bruker spectrometer and was described in previous paper[14]. The fully sampled spectrum consists of 1024×98 complex points, the direct dimension ($^1$H) has 1024 data points while the indirect dimension ($^{15}$N) 98 data points.

The 2D $^1$H–$^{15}$N HSQC spectrum (Fig. S2-4) of GB1 was the data courtesy of Drs. Luke Arbogast and Frank Delaglio at National Institute of Standards and Technology, Institute for Bioscience and Biotechnology Research, USA. The sample was 2 mM U-$^{15}$N, 20%-$^{13}$C GB1 in 25 mM PO4, pH 7.0 with 150 mM NaCl and 5% D$_2$O. Data was collected using a phase-cycle selected HSQC at 298 K on a Bruker Advance 600 MHz spectrometer using a room temp HCN TXI probe, equipped with a z-axis gradient system. The fully sampled spectrum consists of 1676×170 complex points, the direct dimension ($^1$H) has 1676 data points while the indirect dimension ($^{15}$N) 170 data points.

The 2D $^1$H–$^{15}$N best-TROSY spectrum (Fig. S2-5) of ubiquitin was acquired at 298.2K temperature on an 800 MHz Bruker spectrometer and was described in previous paper[14]. The fully sampled spectrum consists of 682×128 complex points, the direct dimension ($^1$H) has 682 data points while the indirect dimension ($^{15}$N) 128 data points.

2.1.2 3D Spectra

The fully sampled 3D HNCO spectrum (Fig. S2-7) of azurin protein obtained from the 800 MHz spectrometer on 15N-13C-labeled Cu(I) azurin sample was described earlier[15]. The fully spectrum is of size 1024×60×60, where its direct dimension ($^1$H) has 1024 points, the indirect dimensions have 60 ($^{15}$N) and 60 ($^{13}$C) points, respectively.

The fully sampled 3D HNCACB spectrum (Fig. S2-10) of GB1-HttNTQ7 was the data courtesy of Drs. Marius Clore and Samuel Kotler at Laboratory of Chemical Physics, National Institute of Diabetes and Digestive and Kidney Diseases, National Institutes of Health, Bethesda, MD 20892-0520. The data was recorded at 298 K on a Bruker Advance HD 700 MHz spectrometer using a cryogenic TCI probe, equipped with a triple-axis gradient accessory, and was described in previous paper[16]. The fully spectrum is of size 1024×90×44, where its direct dimension ($^1$H) has 1024 points, the indirect dimensions have 90 ($^{15}$N) and 44 ($^{13}$C) points, respectively.

The 3D NUS HNCO spectrum (Fig. S2-13) of MALT1 protein was obtained from 0.5 mM 15N/13C/2H-labeled protein in 10 mM Tris pH 7.5, 50 mM NaCl, 1 mM TCEP-d16, 0.002% NaN3, 10 μM DSS-d6 and 10% D2O. NMR experiments were acquired on a Bruker Avance III spectrometer operating at a frequency of 700 MHz for $^1$H using a 5 mm cryo-enhanced inverse resonance QCI HFCN probe at 298 K. More details could be found in the previous paper[17]. Only 30% NUS data were recorded in the experiment. The expected fully spectrum is of size 1024×57×70, where its direct dimension ($^1$H) has 1024 points, the indirect dimensions have 57 ($^{15}$N) and 70 ($^{13}$C) points, respectively.

The 3D NUS HNCO spectrum (Fig. S2-15) of alpha-synuclein was obtained from 0.6mM 15N/13C-labeled alpha-synuclein (20mM Sodium Phosphate buffer, pH=6.5, 200mM NaCl, 0.5mM EDTA, 10% D$_2$O) purchased from Giotto Biotech. This spectrum was recorded at 20°C on an 800MHz Bruker AVANCE III-HD spectrometer equipped with 3mm CP-TCI probe. More details could be found in the previous paper[18]. Only 15% NUS data were recorded in the experiment. The expected fully spectrum is of size 1024×64×64, where its direct dimension ($^1$H) has 1024 points, the indirect dimensions have 64 ($^{15}$N) and 64 ($^{13}$C) points, respectively.

## 2.2 Reconstructed 2D HSQC Spectrum of CD79b

Details about the spectrum could be found in Table S2-1. The deep learning method, DL NMR, is compared with a representative NUS NMR reconstruction method, the LR approach[2].

The DL NMR achieves the same level of reconstructed spectra quality as LR method does (at the NUS rate of 25% in Fig. S2-1). The peak intensity correlation values of both methods approaching 0.9999 and representative peak shapes closing to the fully sampled peak shapes can demonstrate this (Figs. S2-1(d)-(g)). At the lower the NUS levels (10% and 15% in Fig. 2(d)), the DL NMR provides higher correlation values as well as lower dispersion of correlation coefficients over 100 NUS trials. The higher quality of the DL NMR reconstruction at low NUS rate is also illustrated in Figs. S2-2(a) and 2(b). These observations imply that DL allows more significant saving of measurement time than the LR method, and also is more robust under different NUS trials, leading to more stable reconstruction.

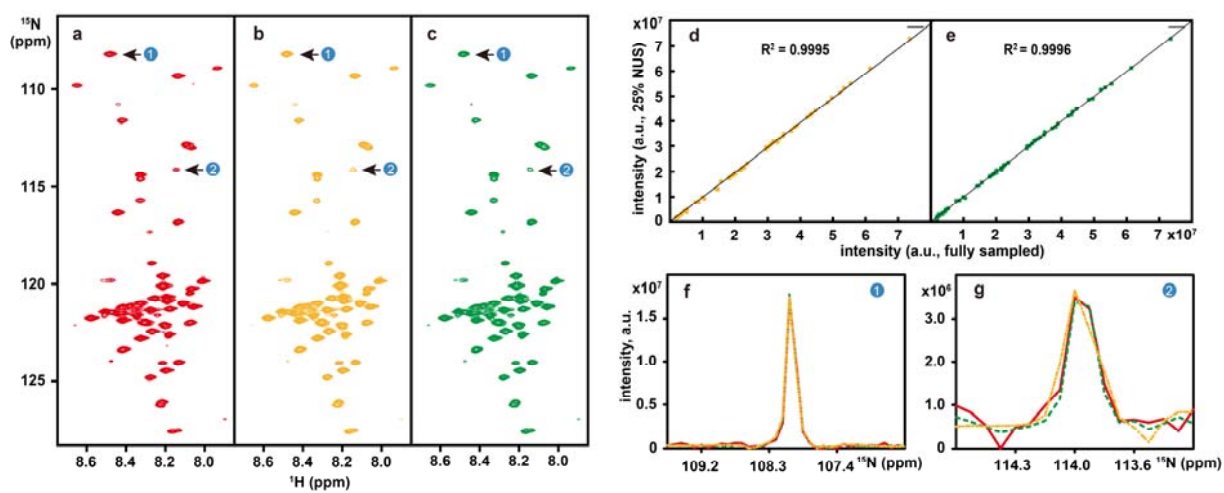

Figure S2-1. Reconstruction of a 2D $^1$H–$^{15}$N HSQC spectrum of cytosolic CD79b protein from the B-cell receptor. (a)-(c) are the fully sampled spectra, LR and DL reconstructions from 25% NUS data, respectively; (d) and (e) are peak intensity correlations obtained by LR and DL methods, respectively; (d) and (e) are zoomed out 1D $^{15}$N traces, and the red, yellow and green lines represent the spectra obtained with fully-sampling, LR and DL methods, respectively. Note: 25% NUS data were used in the reconstruction.

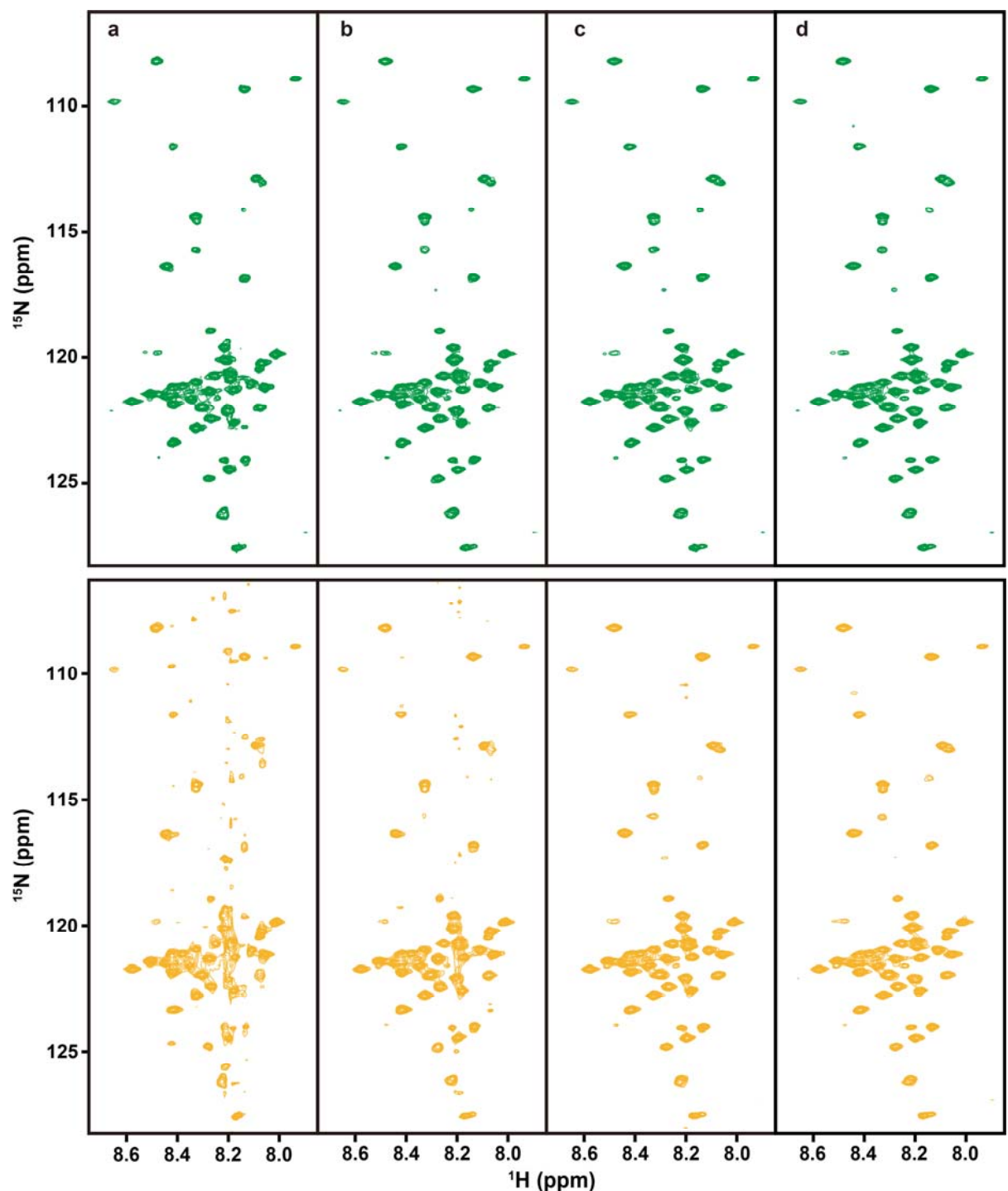

Figure S2-2. Reconstructed 2D $^1$H-$^{15}$N HSQC spectra under different amounts of NUS data. (a)-(d) are the reconstructions at NUS of 10%, 15%, 20% and 25%, respectively. The spectra marked with green and yellow colors are reconstructions with DL and LR methods, respectively.

## 2.3 Other 2D Spectra Reconstruction

To demonstrate the applicability of trained neural networks, we reconstruct another three spectra, including the 2D HSQC spectrum from ubiquitin (Fig. S2-3), the 2D HSQC spectrum from GB1 (Fig. S2-4) and the 2D TROSY spectrum from ubiquitin (Fig. S2-5), details about spectra could be found in Table S2-1.

Both DL and LR methods obtain very high peak intensity correlation (>0.98), which is also confirmed with almost the same peak shapes to the fully sampled spectra (at the NUS rate of 25%). With fewer data, indicating higher acceleration factors of data acquisition, Fig. S2-6 shows that DL outperforms LR in terms of higher intensity correlations.

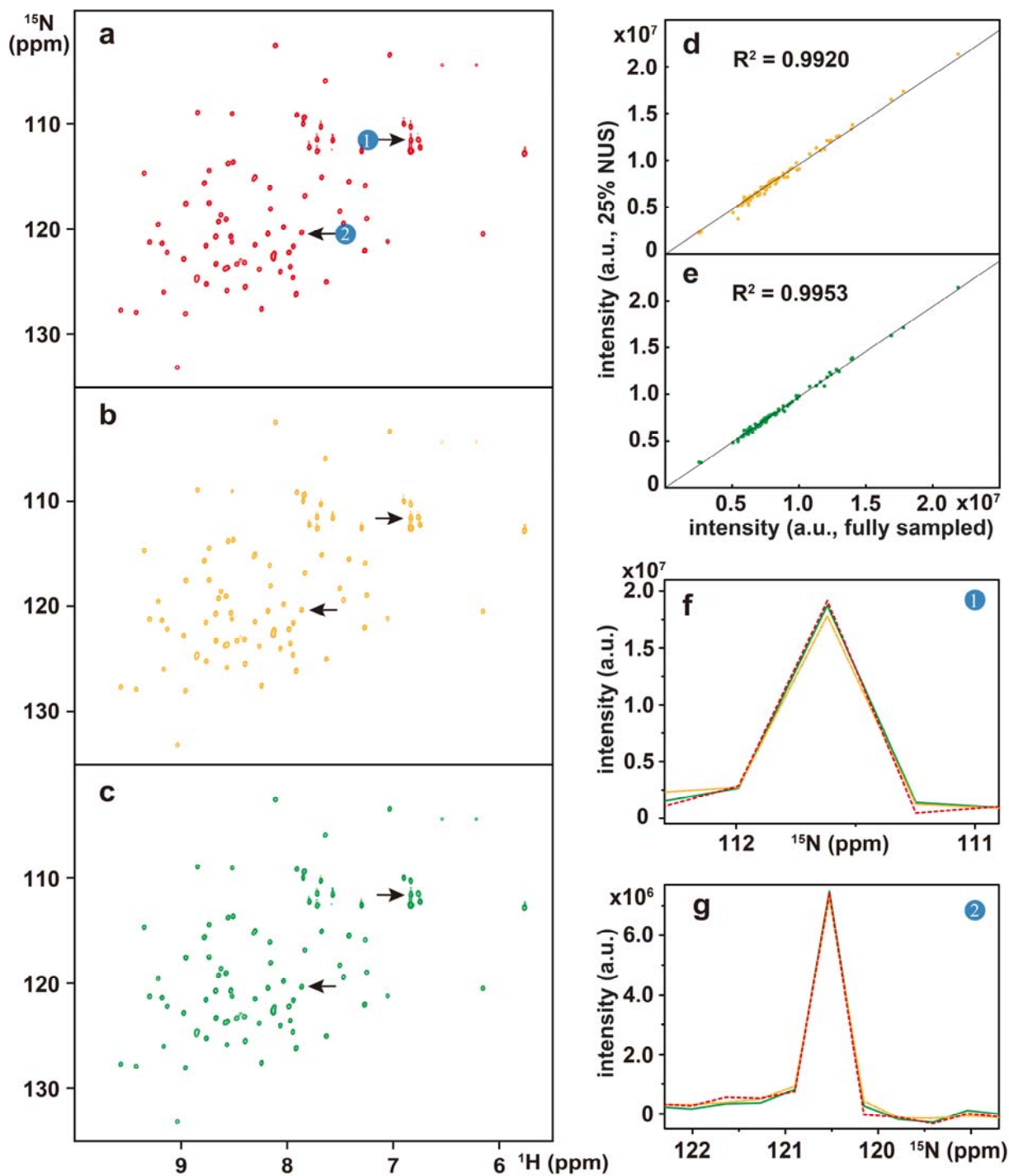

Figure S2-3. Reconstruction of the 2D $^1$H-$^{15}$N HSQC spectrum of ubiquitin. (a) is the fully sampled reference spectrum, (b) and (c) are reconstructed spectra from 25% NUS data by LR and DL methods, respectively, (d) and (e) are the peak intensity correlations achieved by LR and DL methods, respectively, (f) and (g) are zoomed out 1D $^{15}$N traces, and the red, yellow and green lines represent the reference, LR and DL reconstructed spectra, respectively.

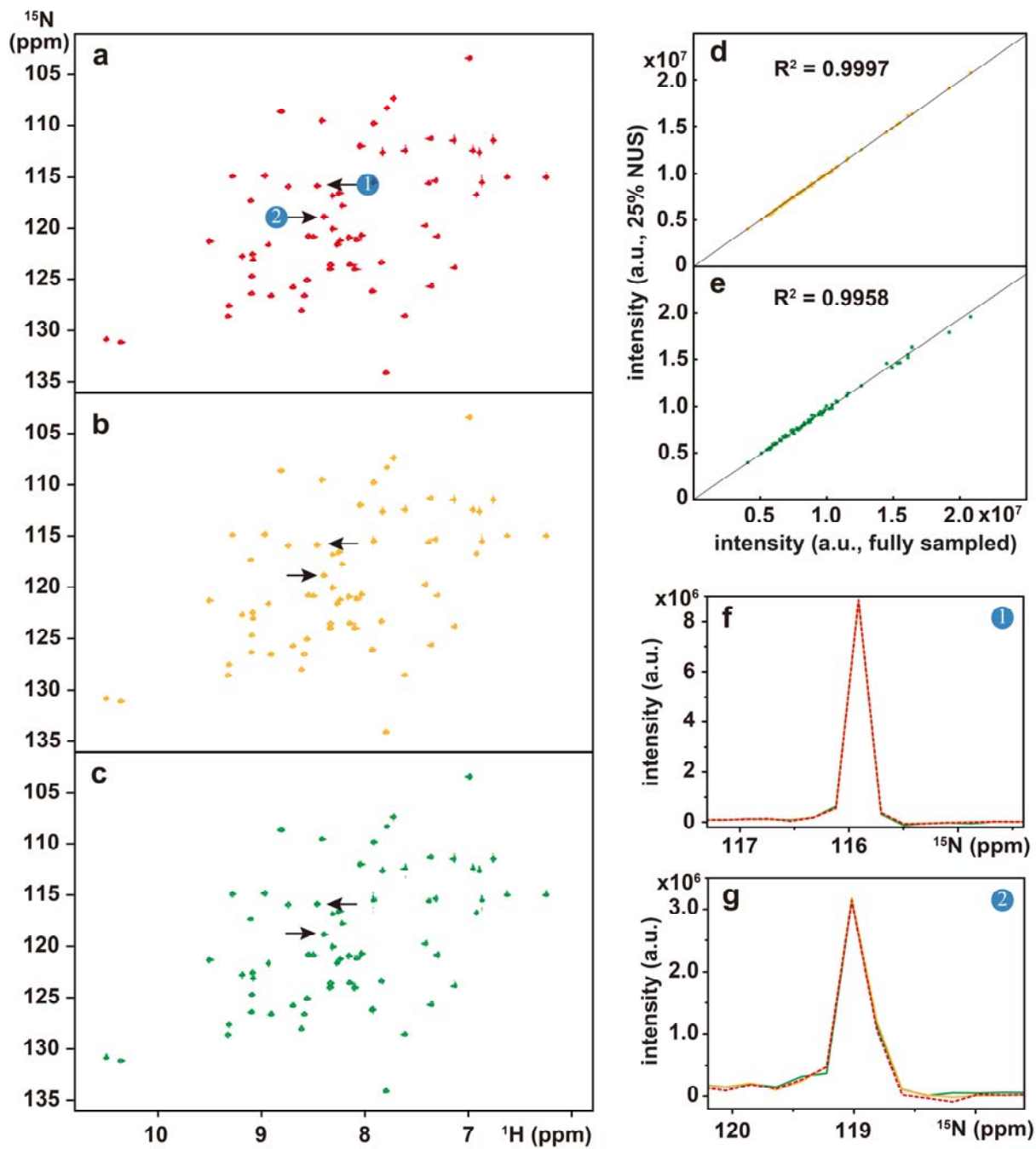

Figure S2-4. Reconstruction of 2D HSQC spectra of GB1. (a) is the fully sampled reference spectrum, (b) and (c) are reconstructed spectra from 25% NUS data by LR and DL methods, respectively, (d) and (e) are the peak intensity correlations achieved by LR and DL methods, respectively, (f) and (g) are zoomed out 1D $^{15}$N traces, and the red, yellow and green lines represent the reference, LR and DL reconstructed spectra, respectively.

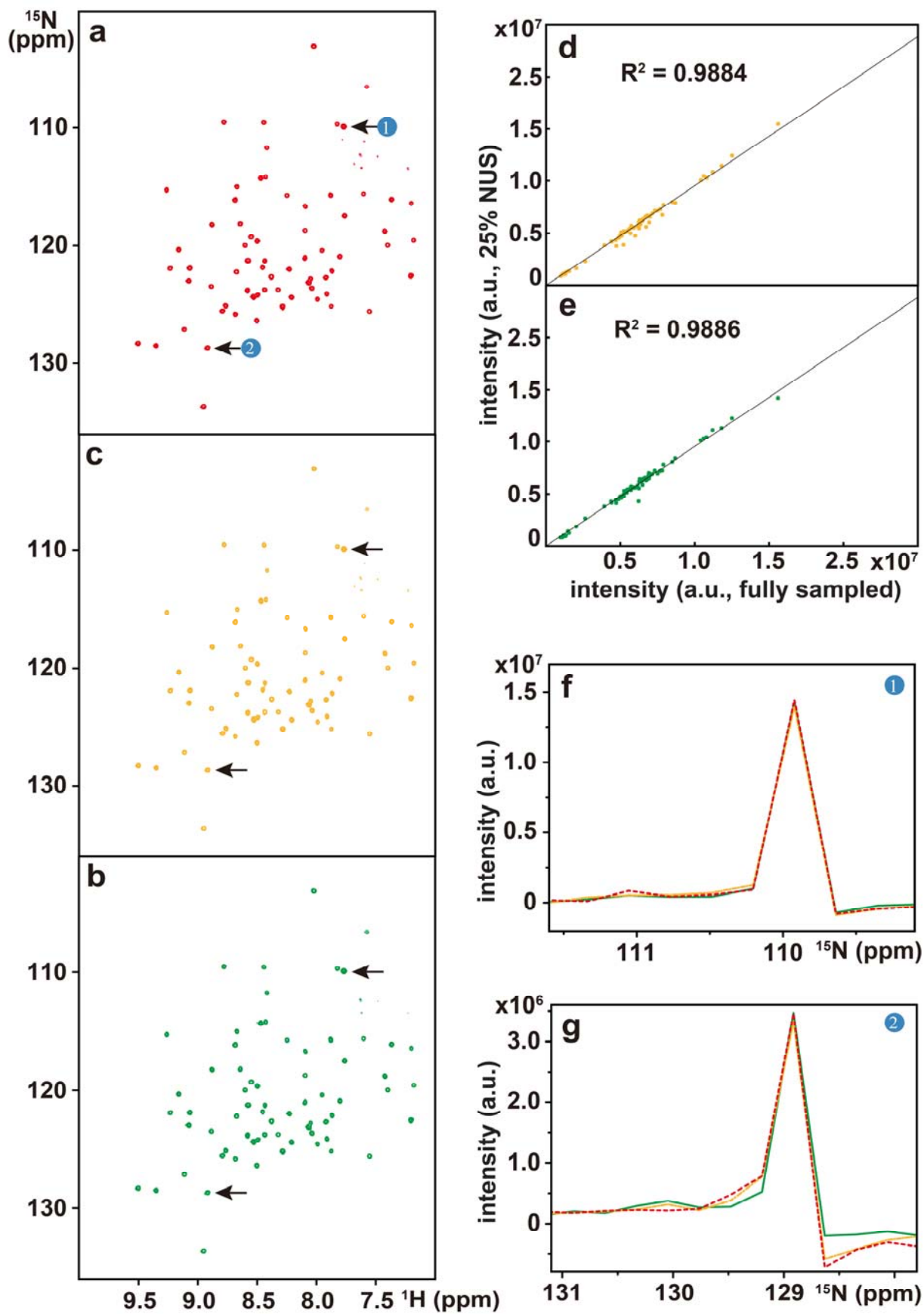

Figure S2-5. Reconstruction of the 2D $^1$H-$^{15}$N best-TROSY spectrum of ubiquitin. (a) is the fully sampled reference spectrum, (b) and (c) are reconstructed spectra from 25% NUS data by LR and deep NMR, respectively,

(d) and (e) are the peak intensity correlations achieved by LR and DL methods, respectively, (f) and (g) are zoomed out 1D $^{15}$N traces, and the red, yellow and green lines represent the reference, LR and DL NMR reconstructed spectra, respectively.

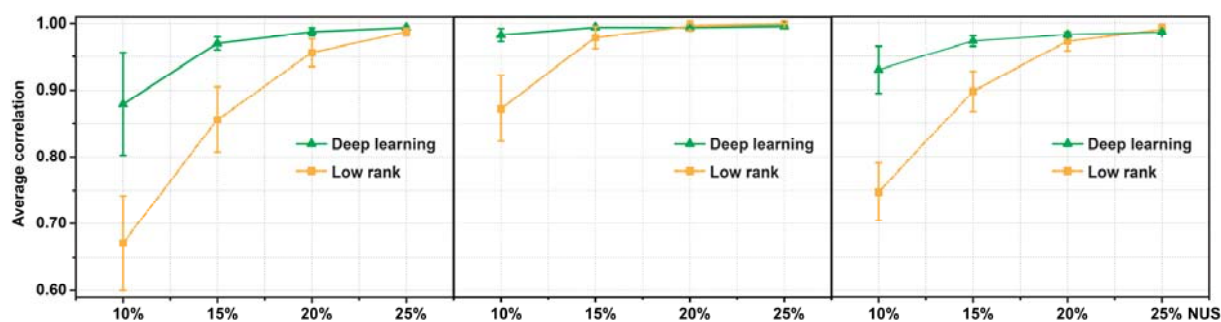

Figure S2-6. Correlation coefficients for the (a) Fig. S2-3, (b) Fig. S2-4 and (c) Fig. S2-5 spectra at different rates of NUS. Note: The green and yellow lines indicate the Pearson correlation coefficient $R^2$ of DL and LR methods, each compared with the fully sampled spectrum, respectively. The error bars are the standard deviations of the correlations over 100 NUS resampling trials.

### 2.4 3D Spectra Reconstruction

In this section, we will demonstrate the applicability of the DL method to 3D NMR spectra reconstruction for small, large and intrinsically disordered protein. The state-of-the-art CS[10] method was adopted for comparison.

The NMR data of MALT1 (44 kDa) and alpha-synuclein (an intrinsically disordered protein, 14.5 kDa) proteins were recorded with non-uniform sampling technique in spectrometers for reducing data acquisition time, while azurin (14 kDa) and GB1-HttNTQ7 (10 kDa) proteins spectra were fully sampled in the experiments. Detail experimental descriptions of these experiments were listed in Table S2-1.

2.4.1 Spectra Reconstruction for Small Proteins

Fully sampled FID data were acquired for two small proteins, including the Azurin (14 kDa) and GB1-HttNTQ7 (10 kDa). The existence of fully sampled spectra would be helpful serving as the golden standard in reconstruction validation. The undersampled FID were obtained by retrospectively undersampling the fully sampled FID and reconstructed by both CS and DL.

As can be seen in Figs. S2-7, S2-8, S2-10 and S2-11, both DL and CS approaches produces nice reconstructions of 3D spectra that are very closing to the fully sampled ones. The peak intensity correlations of DL and CS, with $R^2 > 0.99$, shows the high fidelity of reconstruction (Figs. S2-9 and S2-12).

2.4.2 Spectra Reconstruction for A Large Protein and An Intrinsically Disordered Protein

Partial FID data were acquired for a large protein (MALT1, 44 kDa) and an intrinsically disordered protein (alpha-synuclein, 14.5 kDa). These two spectra were experimentally recorded with non-uniform sampling technique in spectrometers for reducing data acquisition time.

For the large protein, the MALT1, the reconstructed spectra were depicted in Figs. S2-13 and S2-14. Results show that both CS and DL reconstruct the HNCO spectra very well with 30% NUS data, implying that 30% data would be adequate for reconstruction methods to provide reliable results. Here, we performed retrospectively under sampling on the 30% NUS data, taking one out the three data points randomly for emulating the 10% NUS in experiments. The reconstructed spectra by DL (Figs. S2-13(d) & (h)) from 10% NUS data are very similar to the spectra by DL (Figs. S2-13 (b) & (f) and Figs. S2-14(b) & (f)) using 30% NUS data, indicating that DL still offers nice reconstruction even with a very small fraction of data.

For an intrinsically disordered protein, the alpha-synuclein, the reconstructed spectra were shown in Figs. S2-15 and S2-16. Results show that both CS and DL gave nice reconstructions with 15% NUS data (the first two columns in Figs. S2-13 and S2-14). Even under a higher acceleration, only 10% data used for reconstruction, both approaches still allow good reconstruction.

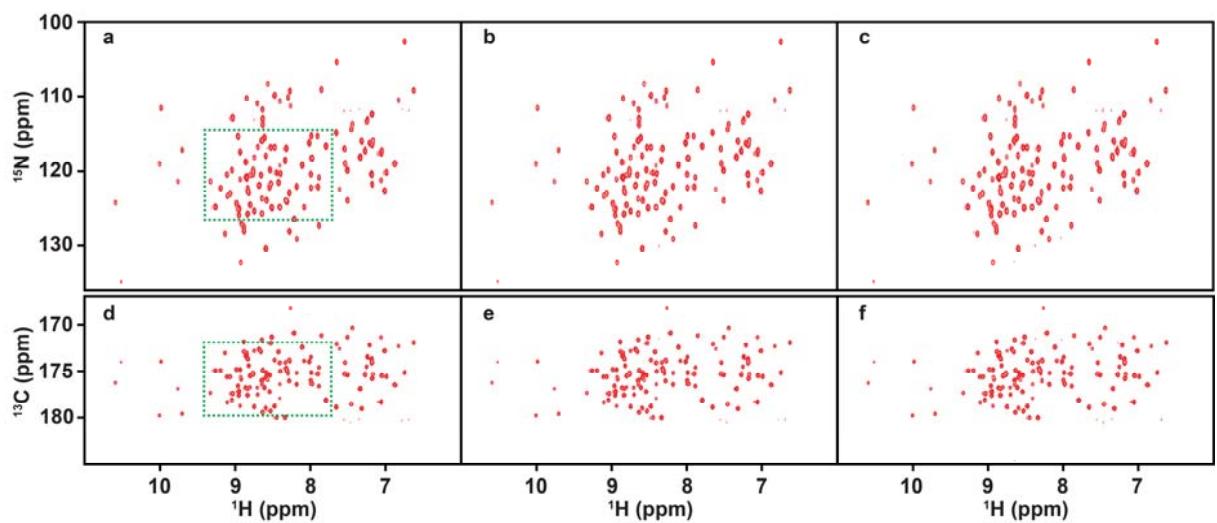

Figure S2-7. The projections on $^1$H-$^{15}$N and $^1$H-$^{13}$C planes of the 3D HNCO spectra of azurin protein. (a) and (d) are projection spectra of the fully sampled referenced spectrum. (b) and (e) are projection spectra of the CS reconstructed spectrum. (c) and (f) are projection spectra of the DL reconstructed spectrum. Note: 5% NUS data were acquired for reconstruction. The sub-region of projections marked with green dash rectangle was shown in Figure S2-8 in the main text. The contours of all projection spectra are at the same level.

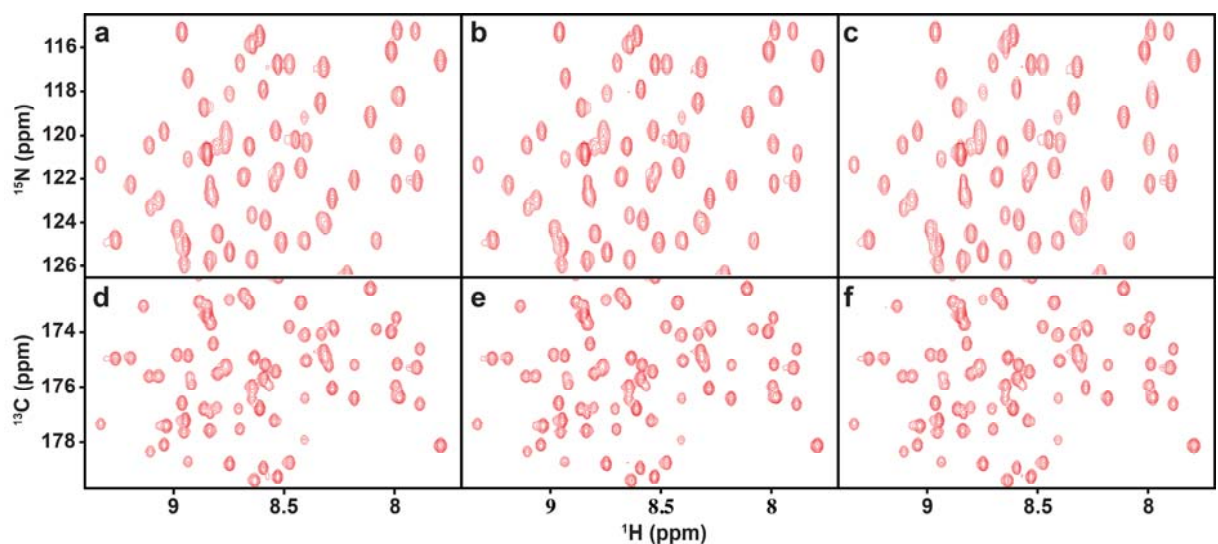

Figure S2-8. The sub-region of the projections on $^1$H-$^{15}$N and $^1$H-$^{13}$C planes of the 3D HNCO of azurin protein. (a) and (d) are projection spectra of the fully sampled referenced spectrum. (b) and (e) are projection spectra of the CS reconstructed spectrum. (c) and (f) are projection spectra of the DL reconstructed spectrum. Note: 5% NUS data were acquired for reconstruction. The contours of all projection spectra are at the same level.

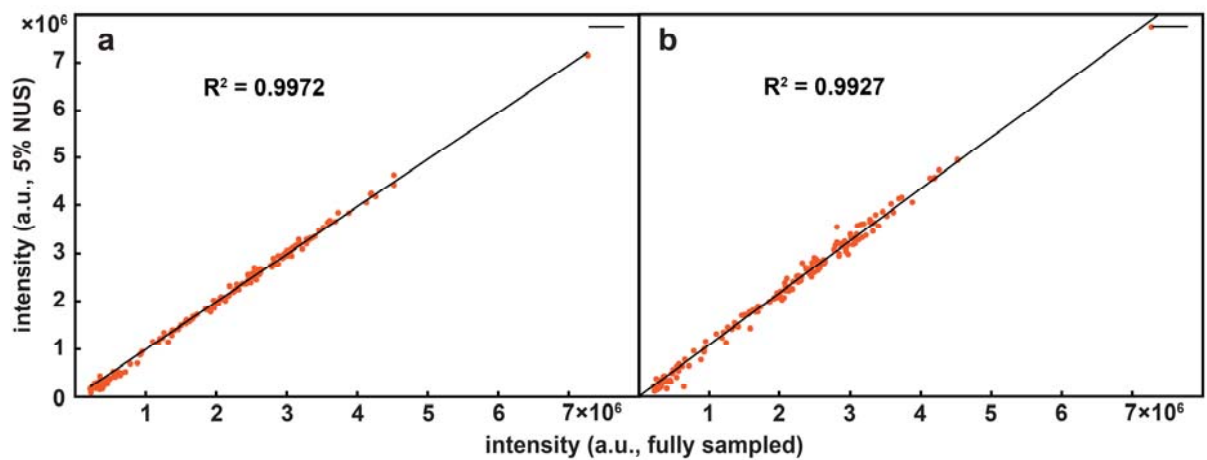

Figure S2-9. Correlation coefficients between reconstructed spectra and fully sampled 3D HNCO shown in Fig. S2-7. (a) and (b) are the peak intensity correlations achieved by CS and DL, respectively.

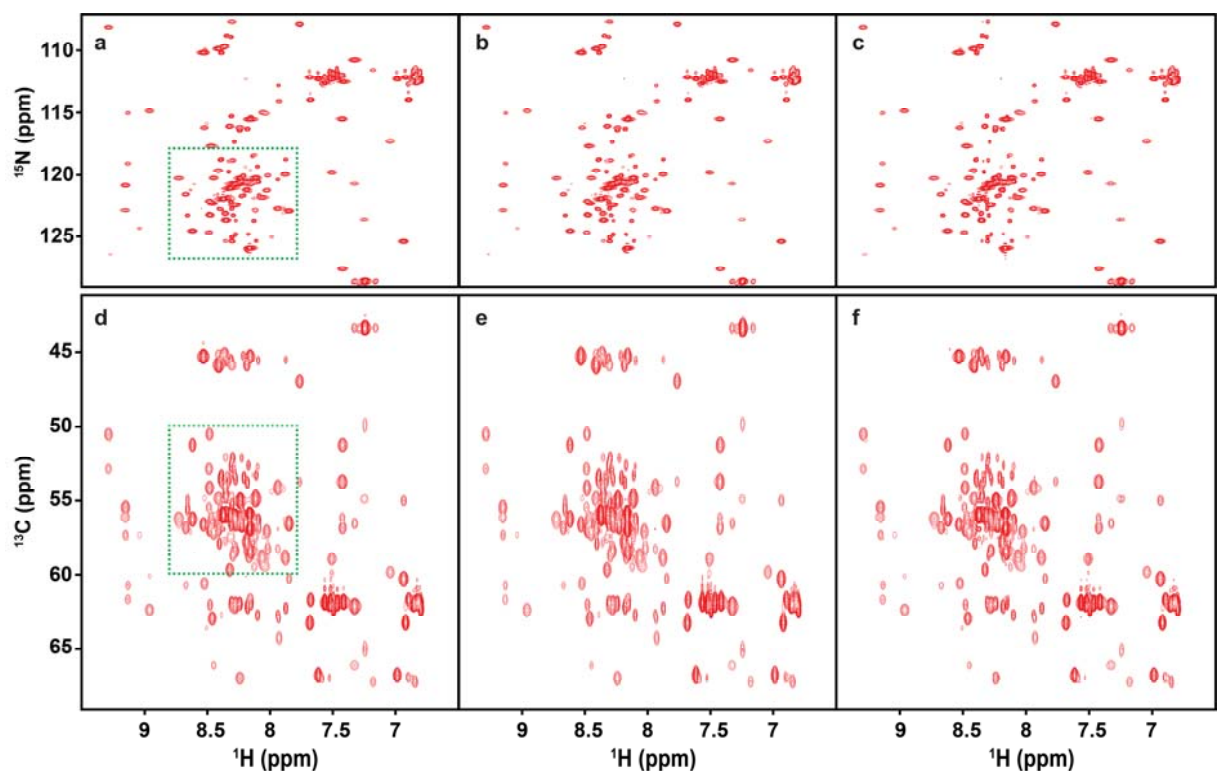

Figure S2-10. The projections on $^1$H-$^{15}$N and $^1$H-$^{13}$C planes of the 3D HNCACB of GB1-HttNTQ7 protein. (a) and (d) are projection spectra of the fully sampled referenced spectrum. (b) and (e) are projection spectra of the CS reconstructed spectrum. (c) and (f) are projection spectra of the DL reconstructed spectrum. Note: 10% NUS data were acquired for reconstruction. The sub-region of projections marked with green dash rectangle was shown in Figure S2-11.

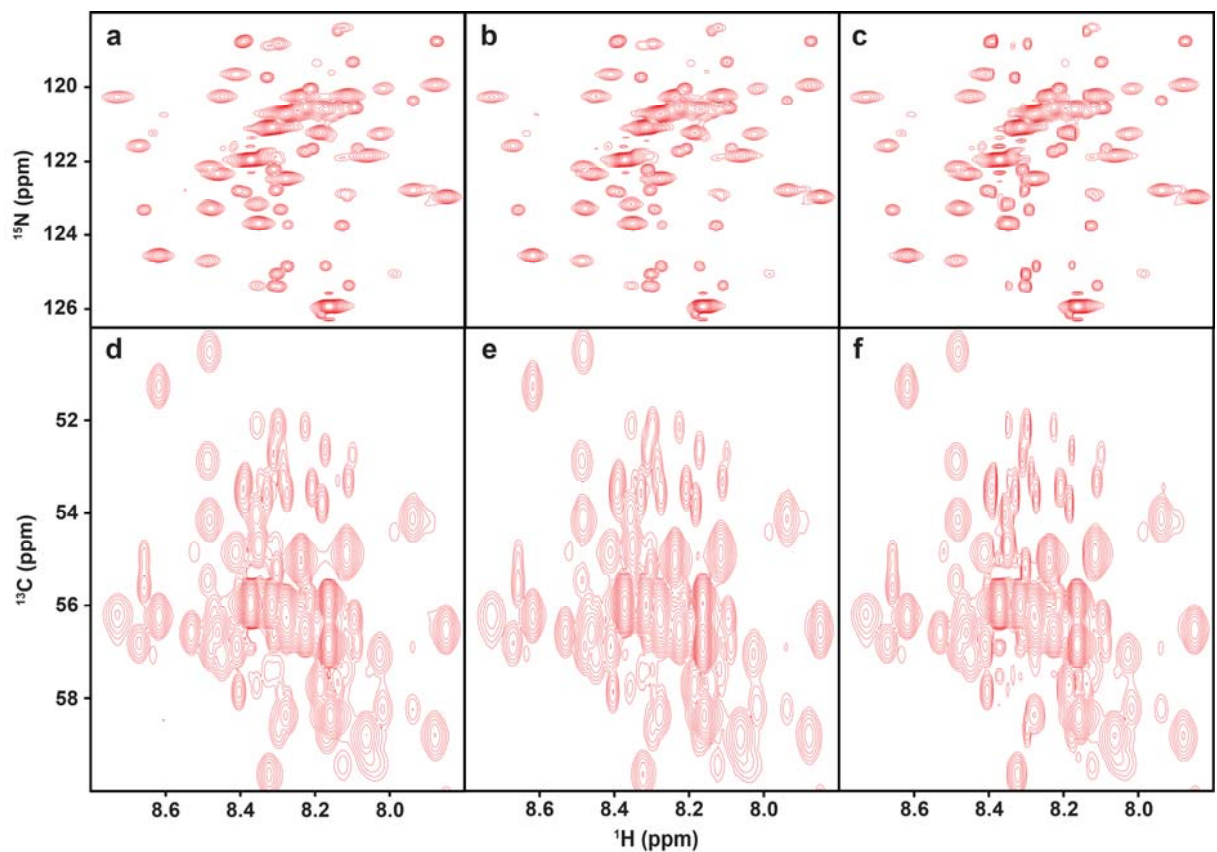

Figure S2-11. The sub-region of the projections on $^1$H-$^{15}$N and $^1$H-$^{13}$C planes of the 3D HNCACB of GB1-HttNTQ7 protein. (a) and (d) are projection spectra of the fully sampled referenced spectrum. (b) and (e) are projection spectra of the CS reconstructed spectrum. (c) and (f) are projection spectra of the DL reconstructed spectrum. Note: 10% NUS data were acquired for reconstruction.

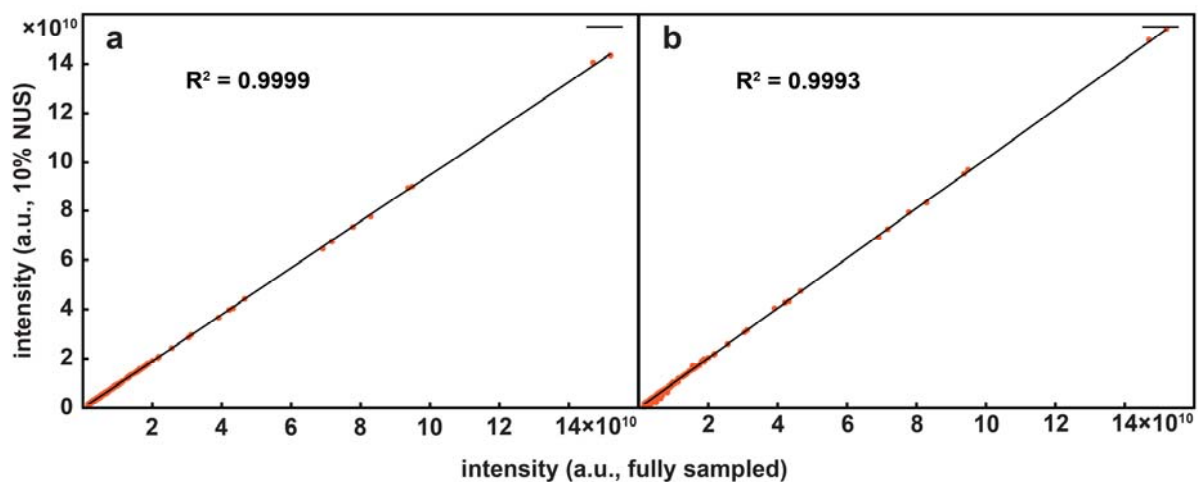

Figure S2-12. Correlation coefficients between reconstructed spectra and fully sampled 3D HNCACB shown in Fig. S2-10. (a) and (b) are the peak intensity correlations achieved by CS and DL, respectively.

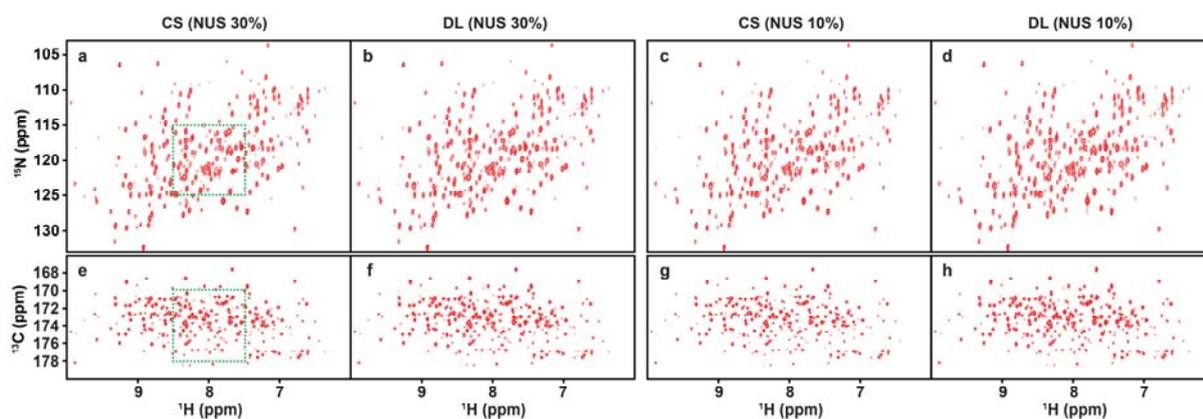

Figure S2-13. The projections on $^1$H-$^{15}$N and $^1$H-$^{13}$C planes of the 3D HNCO spectra of MALT1 protein. (a) and (e) are projection spectra of the CS reconstruction spectrum with 30% NUS data. (b) and (f) are projection spectra of the DL reconstructed spectrum with the same NUS data. (c) and (g) are projection spectra of the CS reconstructed spectrum with 10% NUS data (extracted one out of three points from the 30% NUS data randomly). (d) and (h) are projection spectra of the DL reconstructed spectrum with the same 10% NUS data. Note: The CS reconstructions were scaled by multiplied by a constant for reasonable display. The contours of all projection spectra are at the same level. The sub-region of projections marked with green dash rectangle was shown in Fig. S2-14 and Fig. 4.

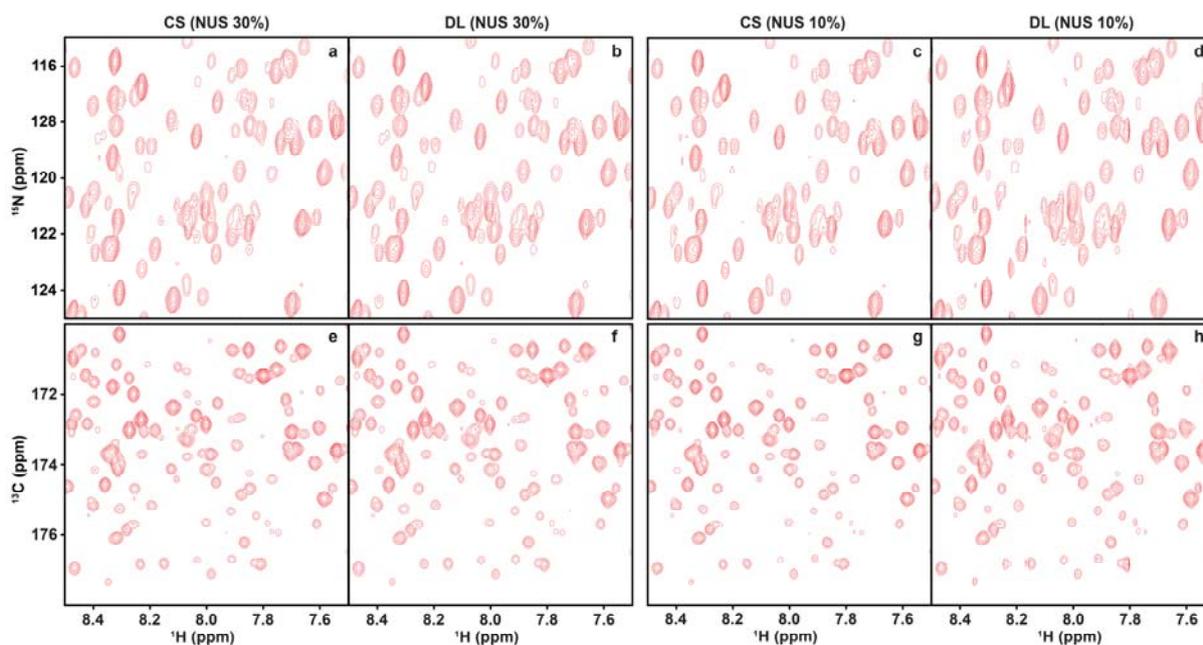

Figure S2-14. The sub-region of the projections on $^1$H-$^{15}$N and $^1$H-$^{13}$C planes of the 3D HNCO spectra of MALT1 protein. (a) and (e) are projection spectra of the CS reconstruction spectrum with 30% NUS data. (b) and (f) are projection spectra of the DL reconstructed spectrum with the same NUS data. (c) and (g) are projection spectra of the CS reconstructed spectrum with 10% NUS data (extracted one out of three points from the 30% NUS data randomly). (d) and (h) are projection spectra of the DL reconstructed spectrum with the same 10% NUS data. Note: The CS reconstructions were scaled by multiplied by a constant for reasonable display. The contours of all projection spectra are at the same level.

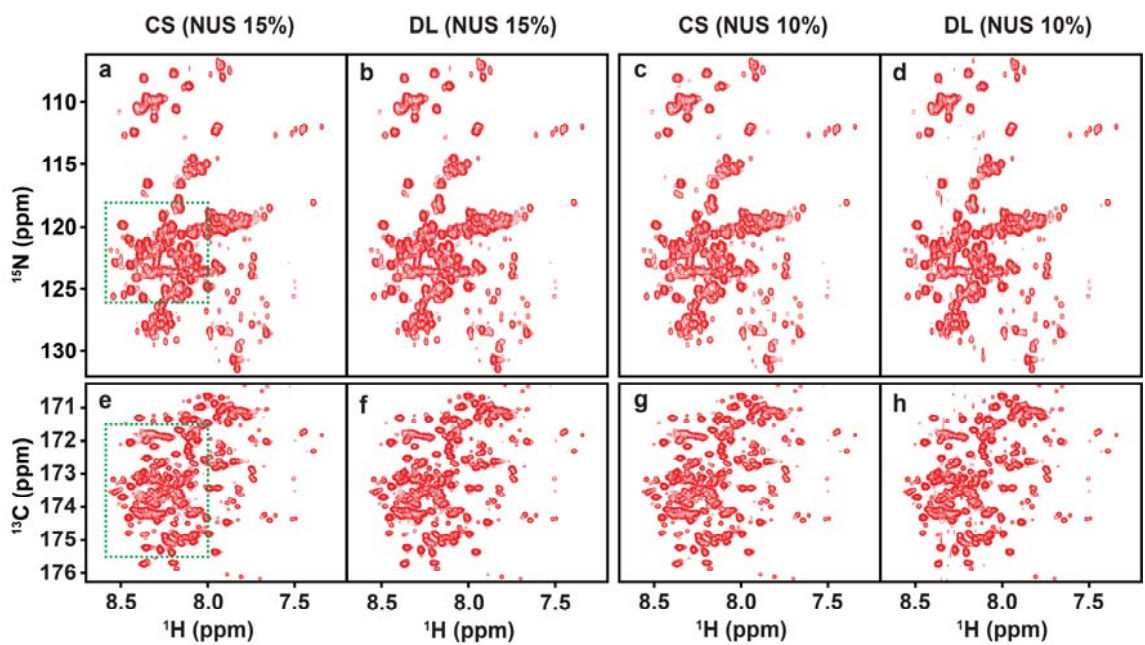

Figure S2-15. The projections on $^1$H-$^{15}$N and $^1$H-$^{13}$C planes of the 3D HNCO spectra of alpha-synuclein protein. (a) and (e) are projections of the CS reconstruction spectrum with 15% NUS data. (b) and (f) are projections of the DL reconstruction with the same NUS data. (c) and (g) are projections of the CS reconstruction with 10% NUS data (extracted two out of three points from the 15% NUS data randomly). (d) and (h) are projections of the DL reconstruction with the same 10% NUS data. Note: The contours of all projection spectra are at the same level. The sub-region of projections marked with green dash rectangle was shown in Fig. S2-16 and Fig. 4.

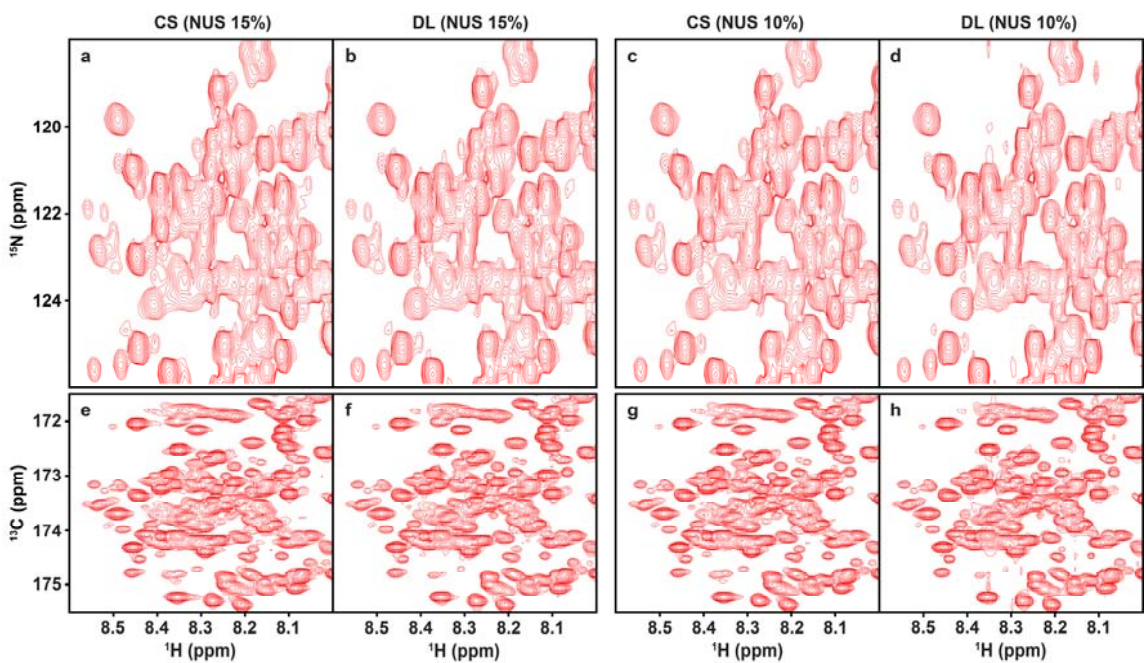

Figure S2-16. The sub-region of the projections on $^1$H-$^{15}$N and $^1$H-$^{13}$C planes of the 3D HNCO spectra of alpha-synuclein protein. (a) and (e) are projection spectra of the CS reconstruction with 15% NUS data. (b) and (f) are projection spectra of the DL reconstruction with the same NUS data. (c) and (g) are projection spectra of the CS reconstruction with 10% NUS data (extracted two out of three points from the 15% NUS data randomly). (d) and (h) are projection spectra of the DL reconstruction with the same 10% NUS data. Note: The contours of all projection spectra are at the same level.

# Supplement S3: Computational Platform and Time

Here, we describe the computational platform and time reported in Fig. 5 of the main text.

All experiments were carried out in a computer server equipped with dual Intel Xeon CPUs (2.2 GHz, 12 cores per CPU), 128 GB RAM, and one Nvidia Tesla K40M GPU card. The proposed deep learning (DL) network was implemented in the open-source deep learning platform – Tensorflow[19], and trained using one Nvidia Tesla K40M GPU card. The low rank (LR) approach was coded in MATLAB[20] and ran with 24 threads. Compressed sensing (CS) with the virtual echo technique[14] in the MddNMR toolbox[10] was ran with 24 threads. Both LR and CS were parallelized with aspects to the direct dimensions to maximally reduce the computation time under multiple CPU cores.

The direct dimensions of all spectra were processed using NMRPipe[12], in which the amide regions were extracted. Then the processed spectra were used for reconstruction by LR[2], CS[10] and the proposed DL NMR. After reconstruction, we also used the NMRPipe[12] to process the indirect dimensions of reconstructed data. The shown spectra were prepared using NMRFAM-SPARK[21]. We compare the time for reconstruction by omitting the time of processing direct and indirect dimensions since those processing are the same for the three methods. The size of spectra to be reconstructed is the dominant factor determining the computational time, thus below we list the size of each direct dimension-processed spectrum, the total runtime time for the whole spectrum and the average time that is calculated by dividing the total runtime by the number of data points in the direct dimension of the spectrum.

Table S3-1. Computational time for 2D spectra (Unit: seconds)

| Type | Protein | Data size ($F_2 \times F_1$) | LR parameters | | | Runtime (LR / DL) | |
|---|---|---|---|---|---|---|---|
| | | | Lambda | Max_iter | Tol | Total time | Average time |
| HSQC | Cytosolic CD79b | 116×256 | 1e5 | 300 | 1e-5 | 1.95 / 0.08 | 0.0168 / 0.0007 |
| HSQC | Ubiquitin | 576×98 | 1e5 | 300 | 1e-5 | 2.68 / 0.17 | 0.0046 / 0.0003 |
| HSQC | Gb1 | 1146×170 | 1e5 | 300 | 1e-5 | 9.95 / 0.63 | 0.0087 / 0.0005 |
| TROSY | Ubiquitin | 512×128 | 1e5 | 300 | 1e-5 | 2.00 / 0.15 | 0.0039 / 0.0003 |

Note: The data size denotes the number of data points to be reconstructed where the first one number is the size of data points in the indirect dimension ($F_2$) and the last number is the size of data points in the direct dimension ($F_1$). In each iteration, LR will evaluate the two convergence criteria, Max_iter (Maximum iteration) and Tol (Tolerance), to determine whether or not to quit the loop. If one of the criteria is met, the program quits.

Table S3-2. Computational time for 3D spectra (Unit: seconds)

| Type | Protein | FID data size ($F_3 \times F_2 \times F_1$) | CS parameters | | Runtime (CS / DL) | |
|---|---|---|---|---|---|---|
| | | | Algorithm | Iteration | Total time | Average time |
| HNCO | Azurin | 732×60×60 | IRLS | 20 | 101.48 / 9.66 | 0.1386 / 0.0132 |
| HNCACB | GB1-HttNTQ7 | 879×90×44 | IRLS | 20 | 72.46 / 13.22 | 0.0824 / 0.0150 |
| HNCO | MALT1 | 735×57×70 | IST | 200 | 137.29 / 11.74 | 0.1868 / 0.0160 |
| HNCO | Alpha-synuclein | 221×64×64 | IST | 200 | 17.11 / 3.84 | 0.0774 / 0.0174 |

Note: The data size denotes the number of data points to be reconstructed where the first number is the size of data points in the direct dimension ($F_3$) and the last two number are the size of data points in the indirect plane ($F_2 \times F_1$).